\documentclass[9pt,twocolumn,twoside,preprint]{osajnl}

\journal{josab2}
\graphicspath{{figures/}}

\usepackage{graphicx}
\usepackage{xcolor}

\usepackage[utf8]{inputenc}
\usepackage[T1]{fontenc}
\usepackage{textcomp}

\usepackage{booktabs}
\usepackage{amsmath,amssymb,amsfonts}
\usepackage{amstext, mathrsfs, textcomp}
\usepackage{mathptmx}
\usepackage{bigints}  
\usepackage{nicefrac}
\usepackage{multirow}
\usepackage{dcolumn}

\usepackage{changes}

\DeclareMathOperator{\Tr}{Tr}

\title{Quantum Theory of Near-field Optical Imaging with Rare-earth Atomic Clusters}

\author[1]{Clément Majorel}
\author[1,*]{Christian Girard}
\author[1]{Aurélien Cuche}
\author[1]{Arnaud Arbouet}
\author[1,2,*]{Peter R. Wiecha}

\affil[1]{CEMES-CNRS, Université de Toulouse, CNRS, Toulouse, France}
\affil[2]{Physics and Astronomy, Faculty of Engineering and Physical Sciences, University of Southampton, Southampton, UK}
\affil[*]{girard@cemes.fr}
\affil[*]{p.wiecha@soton.ac.uk}

\doi{\url{http://dx.doi.org/XX.XXXX/XX.XX.XXXXXX}}

\begin{abstract}
Scanning near-field optical imaging (SNOM) using local active probes provides in general images of the electric part of the photonic local density of states. However, certain atomic clusters can supply more information by simultaneously revealing both the magnetic (m-LDOS) and the electric (e-LDOS) local density of states in the optical range. For example, nanoparticles doped with rare-earth elements like europium or terbium provide  both electric dipolar (ED) and magnetic dipolar (MD) transitions. In this theoretical article, we develop a quantum description of active systems (rare earth ions) coupled to a photonic nanostructure, by solving the optical Bloch equations together with Maxwell's equations. This allows us to access the population of the emitting energy levels for all atoms excited by the incident light, degenerated at the extremity of the tip of a near-field optical microscope. We show that it is possible to describe the collected light intensity due to ED and MD transitions in a scanning configuration. By carrying out simulations on different experimentally interesting systems, we demonstrate that our formalism can be of great value for the interpretation of experimental configurations including various external parameters (laser intensity, polarization and wavelength, the SNOM probe size, the nature of the sample ...).
\end{abstract}

\setboolean{displaycopyright}{true}

\begin{document}

\maketitle

\section{Introduction}
Several experimental methods are able to access the optical Local Density of States (LDOS), shaped by nanostructures integrated in coplanar environments~\cite{girard_near_2005}.
From a historical point of view, it was the local probe method that provided the first images of both electronic \cite{Crommie-Lutz-Eigler:1993,Crampin-Bryant:1996} 
and photonic LDOS \cite{Chicanne-David-Quidant-Weeber-Lacroute-Bourillot-Dereux-Colas-Girard:2002, dereux_subwavelength_2003, Teri-Odom:2009, carminati_electromagnetic_2015, greffet_image_1997}.
In the case of optical experiments, an extremely sharp tip that behaves like a dipolar source of light was brought close to the surface of the sample.
When equipped with a parabolic mirror of large collection angle, such local probes have been used to measure the photonic eigenmodes tailored by the nanostructures on the surface.
Under these specific detection conditions, it was theoretically demonstrated that the signals delivered in this configuration of \textit{scanning near field optical microscope} (SNOM)  are  proportional to the electric part of the photonic LDOS at the scanned positions of the point-like light source~\cite{colas_des_francs_relationship_2001}.
Note that we are leaving aside here electron microscopy tools like electron energy loss spectroscopy (EELS) which are also capable to access the photonic LDOS on a subwavelength spatial resolution~\cite{garcia_de_abajo_optical_2010, arbouet_electron_2014, losquin_link_2015}.

In a second step, this technique has been considerably improved by using 
fluorescing molecules, rare earth atoms, or nitrogen-vacancy (NV) color centers, embedded 
in nanometric beads or crystals, glued at the apex of a sharpened optical fiber \cite{michaelis_optical_2000, cuche_fluorescent_2009, cuche_near-field_2017}.
Dependent on the actual size of the doped nano-crystal, the size of such probe can be viewed as actually point-like, compared to the length-scale of the variations of the optical fields.
A physical interpretation of the mechanism causing the imaging process, has highlighted the role of the \textit{electric} part of the photonic local density of states (e-LDOS) at the emitter position~\cite{desfrancs-girard-dereux:2002}.
Recently, delicate experiments have addressed light emission from rare earth doped emitters, supporting both, strong electric and magnetic dipolar transitions\cite{aigouy_mapping_2014, choi_selective_2016, ernandes_exploring_2018, wiecha_enhancement_2019}. 
While we consider incoherent SNOM with active probes, we note that dielectric resonators have also been proposed recently as coherent SNOM probes to access the magnetic part of the optical near-field \cite{sunProbingVectorialField2018a}.

The main objective of this article is to extend the formalism described in reference \cite{Girard-Martin-Leveque-Colasdesfrancs-Dereux:2005} to the case of dipolar magnetic transitions in order to analyze a large class of experimental configurations where magnetic and electric photo-physical processes are induced by highly confined optical fields in the vicinity of complex environments.
As previously detailed in \cite{Girard-Martin-Leveque-Colasdesfrancs-Dereux:2005}, we  apply a combination of the \textit{Green Dyadic Method} (GDM) with the \textit{optical Bloch equations} used in atomic physics.
The GDM supplies a precise electromagnetic description of the system, by providing maps of both e-LDOS and m-LDOS~\cite{Colas-Girard-Chicane-Peyrade-Weeber-Dereux:2001}. 
These data are then introduced into the Maxwell-Bloch equations to obtain the population evolution of the molecular energy levels and to deduce the atomic emission signals.
Using our extensive formalism for incoherent, quantum-emitter-probe SNOM with electric and magnetic dipole transitions, we discuss several examples involving typical problems and practical SNOM scenarios.

\section{General concepts}
\label{sec:examples}
\begin{figure}[t]
	\centering
	\includegraphics[width=\columnwidth]{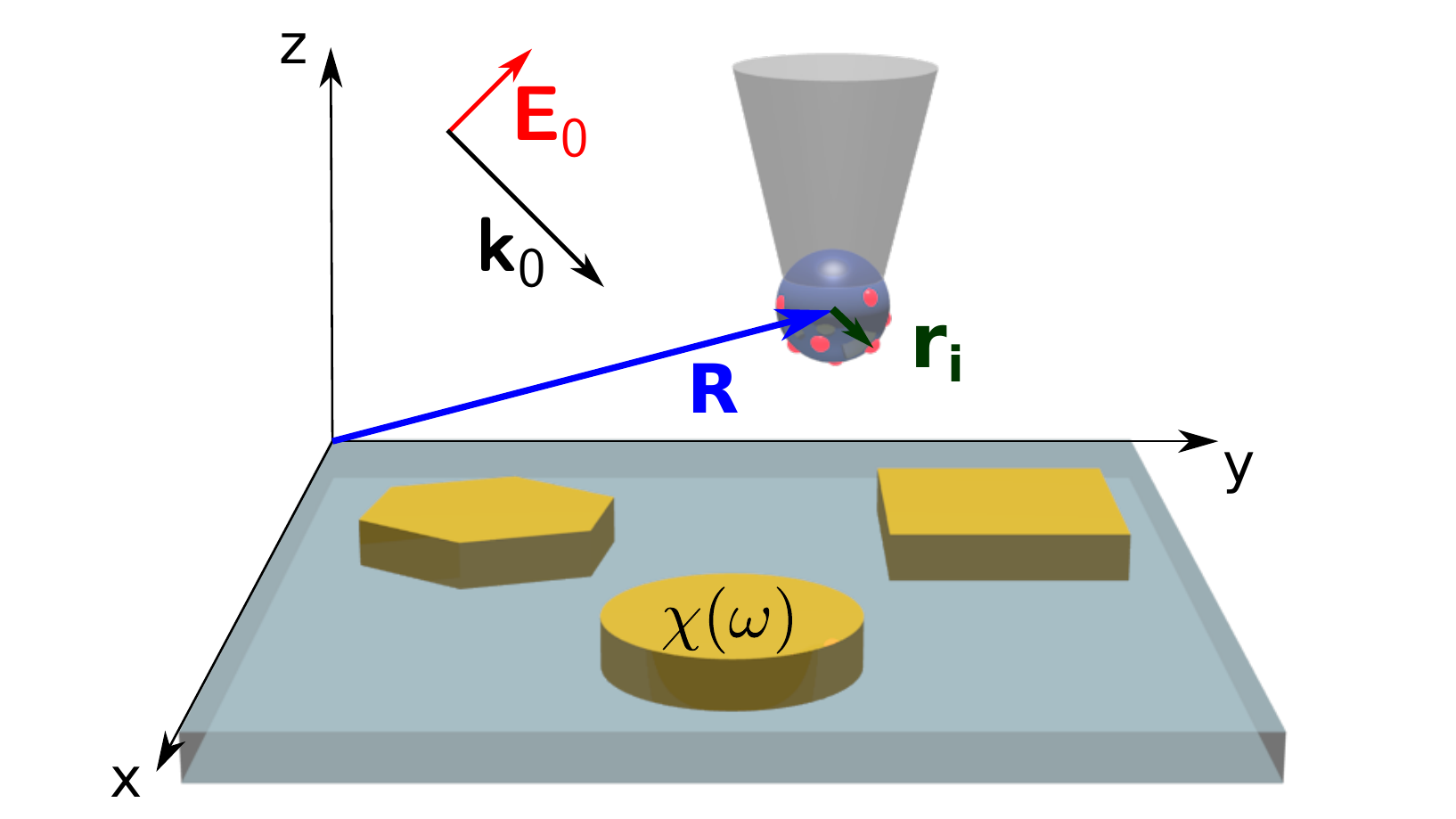}
	\caption{
	General geometry of a quantum near-field microscope in which an europium doped nano-bead raster-scans a transparent sample supporting arbitrary photonic nanostructures.
	The blue arrow represents the location ${\bf R}$ of the tip extremity, the vectors ${\bf r}_{i}$ define the positions of the europium atoms inside the bead, and $\chi (\omega)$ labels the local optical response function of the nanostructures.
	}
\label{fig:overview_quantum_snom}
\end{figure}

As illustrated in figure~\ref{fig:overview_quantum_snom}, in order to establish the main features of our model, we consider a set of $N_{\text{q}}$ quantum systems, located at the positions ${\bf r}_{i}$ from the center $\textbf{R}$ of a dielectric sphere, which is attached to a SNOM tip.

\subsection{Local mean electric field illuminating the quantum systems}

We will discuss other typical configurations later in this article, but for the first demonstration of our approach, we consider an ``apertureless'' SNOM \cite{hillenbrand_pure_2001, huth_nano-ftir_2012, porto_theory_2000}, where the bead containing the quantum structures is illuminated by a monochromatic external plane wave of frequency $\omega_0$, as illustrated in figure~\ref{fig:overview_quantum_snom}.
This excitation field will induce an electric polarization in the nano-bead, which we describe by an isotropic dipolar polarizability $\alpha_{\text{bead}}(\omega)$ via the Clausius–Mossotti relation applied to a spherical volume (see Eq.~(\ref{eq:alpha_calusius_mossotti}) below).
The nanostructures, which are deposited on a dielectric substrate, are consequently illuminated by \(\mathbf{E}_{0}(\mathbf{r},t)\), the superposition of the plane wave and the scattering of the plane wave at the SNOM tip nano-bead.
Finally, the total local field $\mathbf{E}(\mathbf{R},t)$ which drives the emitters in the SNOM tip bead is the superposition of fundamental field and light back-scattered from the nanostructures.
For structures of arbitrary shape and/or materials, $\mathbf{E}(\mathbf{R},t)$ is obtained numerically.
To this end, here we use the Green's Dyadic Method (GDM) to calculate the Fourier transform $\mathbf{E}(\mathbf{R},\omega)$ of $\mathbf{E}(\mathbf{R},t)$.
The GDM is a frequency domain approach based on the idea of a generalized propagator \(\mathbf{K}(\mathbf{r},\mathbf{r}', \omega)\) \cite{martin_generalized_1995}, which contains the response of the nanostructure to any illumination at the frequency \(\omega\):
\begin{equation}\label{eq:generalized_propa_fieldinside}
\mathbf{E}(\mathbf{r}_{\text{source}}, \omega) = \int\limits_{V_{\text{s}}} \mathbf{K}(\mathbf{r}_{\text{source}}, \mathbf{r}', \omega) \cdot \mathbf{E}_{0}(\mathbf{r}', \omega) \text{d}\mathbf{r}'
\end{equation}
where the integral runs over the volume of the nanostructure \(V_{\text{s}}\), which is also called the ``source region'' and \(\mathbf{r}_{\text{source}}\) is a location inside the nanostructure. 
Through a volume discretization of the source region, the integral in equation~(\ref{eq:generalized_propa_fieldinside}) becomes a sum over \(N\) discrete mesh-cells, defining a system of \(3N\) coupled equations~\cite{girard_near_2005}, which can be solved by numerical inversion~\cite{wiecha_pygdmpython_2018}.
After solving equation~(\ref{eq:generalized_propa_fieldinside}), the field inside the source region upon external illumination is known and can be used to calculate the fields at any location outside the nanostructure.
This is done via re-propagation of the field inside the nanostructure using the appropriate Green's tensor \( \mathbf{S}_{\text{env}}(\mathbf{r}, \mathbf{r}', \omega)\) for the chosen reference system (here air \(n_{\text{env}}=1\) above a planar dielectric substrate of index \(n_{\text{s}}=1.5\)):
\begin{multline}\label{eq:field_inside_repropagation}
\mathbf{E}(\mathbf{r}, \omega) = \mathbf{E}_{0}(\mathbf{r}, \omega) \\
 + \int\limits_{V_{\text{s}}} \mathbf{S}_{\text{env}}(\mathbf{r}, \mathbf{r}', \omega) \cdot \chi(\mathbf{r}', \omega) \mathbf{E}(\mathbf{r}', \omega) \text{d}\mathbf{r}'
\end{multline}

\subsection{Near-field coupling with quantum systems}

\begin{figure}[t]
	\centering
	\includegraphics[width=7cm]{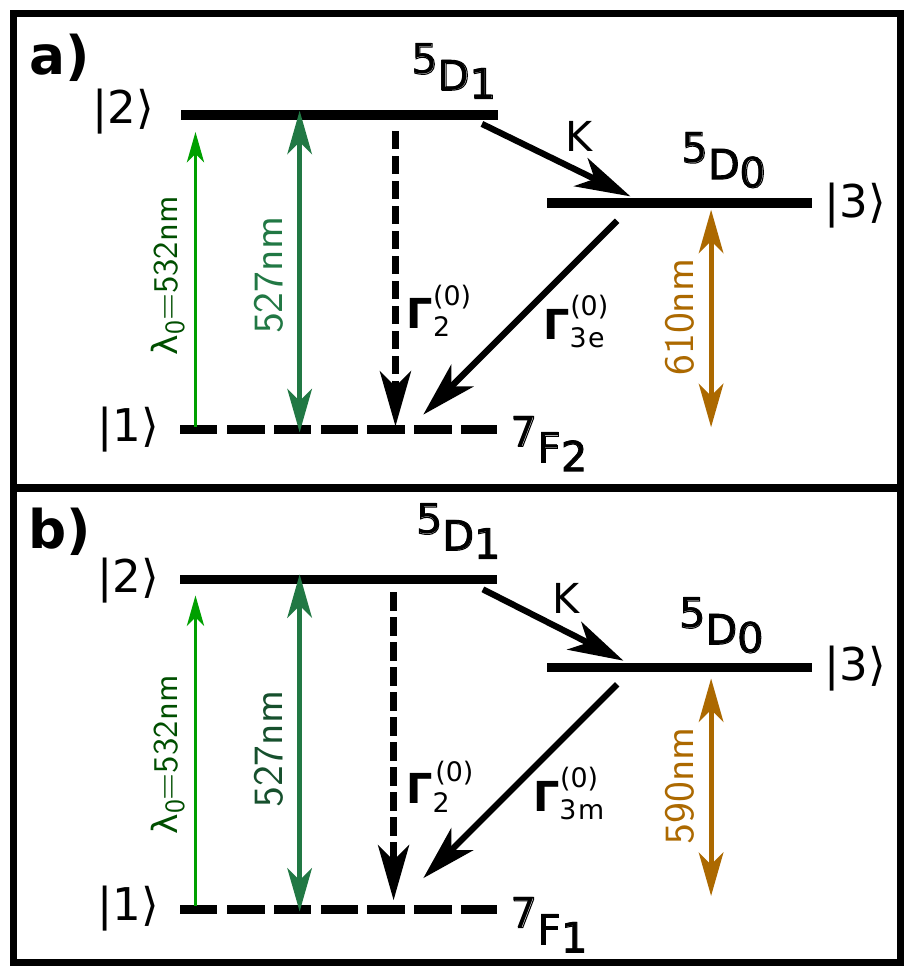}
	\caption{
		Diagram of the Eu\(^{3+}\) ion levels, used to establish the three level quantum system for (a) ED and (b) MD transitions. 
		$\Gamma_{2}^{(0)}$ represents the decay rate between $^{5}D_{1}$ and $^{7}F_{2}$ ($^{7}F_{1}$) states for the ED transition (MD transition), $\Gamma_{3e}^{(0)}$ (MD: $\Gamma_{3m}^{(0)}$) the decay rate between $^{5}D_{0}$ and $^{7}F_{2}$ (MD: $^{7}F_{1}$) and K between $^{5}D_{1}$ and $^{5}D_{0}$.
		We probe the photons emitted by the \(\Gamma_{3e/m}^{(0)}\) transitions.
		The laser wavelength is slightly detuned with respect to the excitation transition energy, to avoid resonant driving. 
		The $^{5}D_{1}$ level can nevertheless be excited due to the finite lifetime \(\Gamma_{2}^{(0)}\).
	}
	\label{fig:europium_levels}
\end{figure}

Our goal is to describe the photo-dynamics in rare-earth atoms, featuring at a time electric dipole (ED) and magnetic dipole (MD) transitions of similar strength.
We will use europium as emitter element throughout this work, but our scheme can be applied likewise to any other ED and/or MD transition, not limited to rare-earth elements.
For both ED and MD transitions the europium photo-dynamics can be described as a three level quantum system \cite{colas_des_francs_theory_2007} (see Fig~\ref{fig:europium_levels}). 
We determine the europium levels which play a role in the photo-dynamics process from \cite{binnemans_interpretation_2015, noginova_magnetic_2008} and use their relative decay rates \(\Gamma_i\) which depend on the environment, as will be further detailed below.
Hence in the following, the level $\vert 1 \rangle$ will correspond to the europium energy level $^{7}F_{2}$ in case of the ED transition, and to $^{7}F_{1}$ in case of the MD transition.
In both, the electric and the magnetic dipole case, $\vert 2 \rangle$ and $\vert 3 \rangle$ correspond to the levels $^{5}D_{1}$ and $^{5}D_{0}$, respectively.
Throughout this article we deliberately choose a slightly higher transition energy between the ground level and $^{5}D_{1}$ (\(\lambda=527\,\)nm) to avoid resonant driving with our \(\lambda_0=532\,\)nm illumination. 
This choice is solely made in order to guarantuee numerical stability of the formalism and has no impact on the physical mechanisms and interpretation.

The below derivation follows essentially previous work \cite{Girard-Martin-Leveque-Colasdesfrancs-Dereux:2005}, which we extend here to ED and MD transitions and whose implications to typical SNOM configurations will be analysed in detail. 
The dipolar coupling Hamiltonian 
\begin{equation}
W(t)=-\boldsymbol{\widehat{\mu}} \cdot\mathbf{E}(\mathbf{R}_{i},t)
\end{equation}
between the local electric field $\mathbf{E}(\mathbf{R}_{i},t)$ and a single 
europium of dipole moment \(\widehat{\mu}\), located at the position $\mathbf{R}_{i} = \mathbf{R} + \mathbf{r}_{i}$ 
relative to the dielectric structure, can be calculated with the GDM (Eq.~(\ref{eq:field_inside_repropagation})).
Using the rotating waves approximation (neglecting non-resonant terms 
in the interaction \cite{cohen-tannoudji_processus_2008}), we then have (with \(k_0=\omega_0/c\))
\begin{equation}\label{eq:4}
W(t)=\sum_{\alpha}\dfrac{\hbar\Omega_{\alpha}(\mathbf{R}_{i},k_{0})}{2}
\left\lbrace e^{\text{i}(\omega_{0}t+\phi_{\alpha}(\mathbf{R}_{i},k_{0}))} \vert 1 \rangle \langle 2 \vert + C.C.\right\rbrace \, .
\end{equation}
The phase factors \(\phi_{\alpha}\) correspond to the phase of the frequency dependent, complex electric field amplitude components \(E_{\alpha}(\mathbf{R}_i, \omega)\) from the field in Eq.~(\ref{eq:field_inside_repropagation}), where \(\alpha\) is one of \(x\), \(y\) or \(z\).
$\Omega_{\alpha}(\mathbf{R}_{i},k_{0}) $ is the $\alpha$-component of the vectorial Rabi frequency correlated to the $\alpha$-component of the absorption transition dipole~$\mu_{12}^{\alpha}$ (i.e. the strength of the emitter excitation): 

\begin{equation}\label{eq:5}
\Omega_{\alpha}(\mathbf{R}_{i},k_{0}) = -\dfrac{\mu_{12}^{\alpha}E_{\alpha}(\mathbf{R}_{i},k_0)}{\hbar}
\end{equation}
where 
\begin{equation}
E_{\alpha}(\mathbf{R}_{i},k_0) = 
\Bigg| \sum\limits_{\beta} \int K_{\alpha\beta}(\mathbf{R}_i, \mathbf{r}', -\omega_0) e^{\text{i} k_0 \mathbf{r}'} \, \text{d}\mathbf{r}' \, E_{0,\beta} \Bigg| \, .
\end{equation}
In the latter equation, \(K_{\alpha\beta}\) is the $\alpha\beta$ component of the generalized propagator \(\mathbf{K}\) given in Eq.~(\ref{eq:generalized_propa_fieldinside}) and summation is performed over \(\beta \in \{x,y,z\}\).

We can now use the Liouville's equation that governs the density matrix evolution of the $i$-th rare earth atom~\cite{cohen-tannoudji_processus_2008}
\begin{equation}\label{eq:6}
\dot{\rho}(t) = \dfrac{1}{\text{i}\hbar} \Big\{ H_{0}(t)+W(t),\,\rho(t)\Big\} + \mathcal{R}_{\text{spont}}\ \rho(t).
\end{equation}
where \(\dot{\rho}(t)={\partial\rho(t)}/{\partial t}\) is the time derivative of the density matrix, \(\{\dots\}\) is the Poisson bracket and the Redfield operator $\mathcal{R}_{\text{spont}}$ describes the coupling with the photon-bath.
In our optical problem, only the populations of the excited levels ($\vert 2 \rangle$ and $\vert 3 \rangle$) need to be considered (see Fig. \ref{fig:europium_levels}). 
Consequently, we focus on this coupled system of four optical Bloch equations~\cite{Girard-Martin-Leveque-Colasdesfrancs-Dereux:2005}
\begin{subequations}\label{eq:7}
\begin{equation}\label{eq:7a}
\dot{\rho}_{33}(t) = K\rho_{22}(t)-\Gamma_{3m/e}\rho_{33}(t)
\end{equation}
\begin{multline}\label{eq:7b}
\dot{\rho}_{22}(t) = -(K+\Gamma_{2})\rho_{22}(t) \\
- \text{i}\left( \Omega^{-}e^{-\text{i}\omega_{0}t}\rho_{12}(t)-\Omega^{+}e^{\text{i}\omega_{0}t}\rho_{21}(t)\right)
\end{multline}
\begin{multline}\label{eq:7c}
\dot{\rho}_{12}(t)  = -\dfrac{\Gamma_{2}}{2}\rho_{12}(t)+\text{i}\omega_{21}\rho_{12}(t) \\ 
- \text{i}\Omega^{+}e^{\text{i}\omega_{0}t}(2\rho_{22}(t)+\rho_{33}(t)-1)
\end{multline}
\begin{multline}\label{eq:7d}
\dot{\rho}_{21}(t) = -\dfrac{\Gamma_{2}}{2}\rho_{21}(t) - \text{i}\omega_{21}\rho_{21}(t) \\
- \text{i}\Omega^{-}e^{-\text{i}\omega_{0}t}(2\rho_{22}(t)+\rho_{33}(t)-1)
\end{multline}
\end{subequations}
with
\begin{equation}\label{eq:8}
\Omega^{\pm}=\sum_{\alpha}\dfrac{\Omega_{\alpha}(\mathbf{R}_{i},\,k_0)}{2}e^{\pm \text{i}\phi_{\alpha}(\mathbf{R}_{i},\,k_0)}
\end{equation}
Now, we will use the approximation that the phase factors $\phi_{\alpha}(\mathbf{R}_{i},\,k_0)$ 
don't vary significantly between the three field components, hence $\phi_{\alpha}(\mathbf{r}_{i},\,k_0)= \phi(\mathbf{r}_{i},\,k_0)$.
According to this approximation $\Omega^{-}=(\Omega^{+})^{*}$.
We also apply a change of variables to remove the phase terms $e^{\pm i\omega_{0}t}$, and we obtain the system
\begin{subequations}\label{eq:9}
\begin{align}
\label{eq:9a}
\dot{\widehat{\rho_{33}}} & = K\widehat{\rho_{22}}-\Gamma_{3m/e}\widehat{\rho_{33}} \\ 
\label{eq:9b}
\dot{\widehat{\rho_{22}}} & = -(K+\Gamma_{2})\widehat{\rho_{22}}-\text{i}\Omega^{-}\widehat{\rho_{12}}+\text{i}\Omega^{+}\widehat{\rho_{21}} \\ 
\label{eq:9c}
\dot{\widehat{\rho_{12}}} & = \text{i}\delta_{L}\widehat{\rho_{12}}-\dfrac{\Gamma_{2}}{2}\widehat{\rho_{12}} -\text{i}\Omega^{+}(\widehat{\rho_{22}}-\widehat{\rho_{11}})\\
\label{eq:9d}
\dot{\widehat{\rho_{21}}} & = -\text{i}\delta_{L}\widehat{\rho_{21}}-\dfrac{\Gamma_{2}}{2}\widehat{\rho_{21}} +\text{i}\Omega^{-}(\widehat{\rho_{22}}-\widehat{\rho_{11}})
\end{align}
\end{subequations}
where $\delta_{L}=\omega_{21}-\omega_{0}$ represents the detuning in frequency between the incident field and the resonant absorption transition between the $^{7}F$ levels and $^{5}D_{1}$.
K is the relaxation constant between $^{5}D_{1}$ and $^{5}D_{0}$, $\Gamma_{2}$ and $\Gamma_{3m/e}$ the two radiative decay rates.

We now place ourselves in the stationary regime where all the temporal derivatives equal zero. 
After solving the system of coupled equations, we obtain the population of level $\vert 3 \rangle$
\begin{equation}\label{eq:10}
\widehat{\rho_{33}}\left(\mathbf{R}_{i}\right) = \dfrac{1}
{1+\left( A_{1}+A_{2}\left(\mathbf{R}_{i}\right)+A_{3}\left(\mathbf{R}_{i}\right)\right) \Gamma_{3m/e}\left(\mathbf{R}_{i}\right)}
\end{equation}
with
\begin{subequations}\label{eq:11}
\begin{align}
\label{eq:11a}
&A_{1}=\dfrac{2}{K}\, ,\\  
\label{eq:11b}
&A_{2}\left(\mathbf{R}_{i}\right)=\dfrac{\delta_{L}^{2}}{\Omega^{+}\left(\mathbf{R}_{i}\right)	\Omega^{-}\left(\mathbf{R}_{i}\right)\Gamma_{2}}+\dfrac{\delta_{L}^{2}}{\Omega^{+}\left(\mathbf{R}_{i}\right)\Omega^{-}\left(\mathbf{R}_{i}\right)K}\, ,\\  
\label{eq:11c}
&A_{3}\left(\mathbf{R}_{i}\right)=\dfrac{\Gamma_{2}}{4\Omega^{+}\left(\mathbf{R}_{i}\right)\Omega^{-}\left(\mathbf{R}_{i}\right)}+\dfrac{\Gamma_{2}^{2}}{4\Omega^{+}\left(\mathbf{R}_{i}\right)	\Omega^{-}\left(\mathbf{R}_{i}\right)K} \,.
\end{align}
\end{subequations}
For a single europium located at the position $\mathbf{R}_{i}$, the fluorescence signal emitted by this system is \cite{Girard-Martin-Leveque-Colasdesfrancs-Dereux:2005}
\begin{equation}
\begin{aligned}\label{eq:12}
I^{m/e}\left(\mathbf{R}_{i}\right) 
 =  \hbar\, \omega_{31}\, \widehat{\rho_{33}}\left(\mathbf{R}_{i}\right) \Gamma_{3m/e}(\mathbf{R}_{i}) \, .
\end{aligned}
\end{equation}
To consider the emission of all europium atoms at the apex of the SNOM tip, we will neglect any optical coupling between the emitters. 
In this case, the fluorescence signal of the entire quantum system is simply the sum of the individual europium emitters
\begin{equation}\label{eq:13}
I^{m/e}_{tot} = \sum\limits_{i=1}^{N_{\text{q}}} I^{m/e}(\mathbf{R}_i) \, ,
\end{equation}
with $N_{\text{q}}$ the number of europium atoms embedded in the dielectric nano-bead. 
For a small tip $\mathbf{r}_{i}\ll \mathbf{R}$, the total signal can be approximated by \(I^{m/e}_{tot} \approx N_{\text{q}} \, I^{m/e}(\mathbf{R})\).

\subsection{Electric and magnetic LDOS of photonic nanostructures}

For the evaluation of the fluorescence signal equation~(\ref{eq:12}), we need to know the (relative) decay rates for all optical transitions involved in the excitation / emission process (see Fig.~\ref{fig:europium_levels}).
To demonstrate the general approach to calculate these decay rates, we consider the vacuum decay rates associated with the MD ($\Gamma_{3m}^{(0)}$) and ED transition ($\Gamma_{3e}^{(0)}$), depicted in figure~\ref{fig:europium_levels}. 
These decay rates are modified in presence of a photonic nanostructure~\cite{agarwal_quantum_1975, kwadrin_probing_2013, baranov_modifying_2017, wiecha_decay_2018, rolly_promoting_2012, lunnemann_local_2016}:
\begin{equation}\label{eq:14}
\Gamma_{3m/e}(\mathbf{R}_{i})=\dfrac{2\left(\mu^{m/e}_{31}\right)^{2}}{\hbar}\text{Im}\left\lbrace \mathbf{S}^{HH/EE}(\mathbf{R}_{i},\mathbf{R}_{i},\omega_{31}) :\mathbf{u}\mathbf{u} \right\rbrace 
\end{equation}
where $\mu_{31}^{m}$ ($\mu_{31}^{e}$) is the magnetic (electric) transition dipole amplitude, $\omega_{31}$ is the frequency associated with the transition and $\mathbf{u}$ is the unit vector along its orientation.
$\mathbf{S}^{HH}$ is the magnetic field susceptibility tensor associated with the MD transition of the europium and $\mathbf{S}^{EE}$ corresponds to the electric field susceptibility tensor related to the ED transition.
Whatever the nature of the transition, the field susceptibility can be developed as the sum of two tensors 
\begin{equation}\label{eq:15}
\mathbf{S}^{HH/EE}(\mathbf{R}_{i},\mathbf{R}_{i},\omega_{31})=\mathbf{S}_{0}^{HH/EE}(\mathbf{R}_{i},\mathbf{R}_{i},\omega_{31})+\mathbf{S}_{p}^{HH/EE}(\mathbf{R}_{i},\mathbf{R}_{i},\omega_{31})
\end{equation}
where the tensor $\mathbf{S}_{0}^{HH/EE}(\mathbf{R}_{i},\mathbf{R}_{i},\omega_{31})$ represents the response function of vacuum (in general this can be replaced by a field susceptibility describing the environment, including the substrate, without perturbation) and $\mathbf{S}_{p}^{HH/EE}(\mathbf{R}_{i},\mathbf{R}_{i},\omega_{31})$ the response associated to the nanostructure.
In case of the magnetic decay rate, replacing the $\mathbf{S}^{HH}(\mathbf{R}_{i},\mathbf{R}_{i},\omega_{31})$ term in Eq.~(\ref{eq:14}) by its developed expression (Eq.~\ref{eq:15}) leads to \cite{wiecha_decay_2018} 
\begin{equation}\label{eq:16}
\Gamma_{3m}(\mathbf{R}_{i})=\Gamma^{(0)}_{3m}+\dfrac{2\left(\mu^{m}_{31}\right)^{2}}{\hbar}\text{Im}\left\lbrace \mathbf{S}_{p}^{HH}(\mathbf{R}_{i},\mathbf{R}_{i},\omega_{31}):\mathbf{u}\mathbf{u}\right\rbrace \, ,
\end{equation}
where the superscript ``$(0)$'' in $\Gamma^{(0)}_{3m}$ refers to the spontaneous decay rate of the isolated emitter(s), without the presence of the perturbation.
In the electric case, we have a similar expression of the decay rate \cite{Girard-Martin-Leveque-Colasdesfrancs-Dereux:2005}. 
Now, if the dipole is along a specific direction $\alpha \in \{ x,\, y,\, z\}$, we can reformulate the equation (\ref{eq:16}) by introducing the partial, magnetic photonic LDOS $n^{H}_{\alpha}(\mathbf{R}_{i},\omega_{31})$ \cite{Colas-Girard-Chicane-Peyrade-Weeber-Dereux:2001}
\begin{equation}\label{eq:17}
n^{H}_{\alpha}(\mathbf{R}_{i},\omega_{31})=\dfrac{1}{2\pi^{2}\omega_{31}}\text{Im}\Big\lbrace \mathbf{S}^{HH}_{p_{\alpha\alpha}}(\mathbf{R}_{i},\mathbf{R}_{i},\omega_{31})\Big\rbrace 
\end{equation}
Finally, if we introduce Eq. (\ref{eq:17}) in Eq. (\ref{eq:16}) we obtain the partial decay rate for an emitter with orientation along direction~\(\alpha\)
\begin{equation}\label{eq:18}
\Gamma_{3m}^{\alpha}(\mathbf{r_{m}}) = 
\Gamma^{(0)}_{3m} + \dfrac{4\pi^{2}\omega_{31}\left(\mu^{m}_{31}\right)^{2}}{\hbar}n^{H}_{\alpha}(\mathbf{R}_{i},\omega_{31})\, .
\end{equation} 
Analogously the LDOS can be calculated for an ED transition.

\subsection{A simple analytical model}

The above introduced partial LDOS and decay rates for the electric and magnetic transitions can be obtained using a volume discretization scheme~\cite{wiecha_decay_2018, wiecha_pygdmpython_2018} for arbitrarily oriented emitters and for arbitrary illumination conditions on arbitrary nanostructures.
Before we analyze the general case for several typical SNOM configurations, we first want to assess the behavior of the signal from the europium doped SNOM tip using a simplified, analytical model example.

\begin{figure}[t]
	\centering
	\includegraphics[width=\columnwidth]{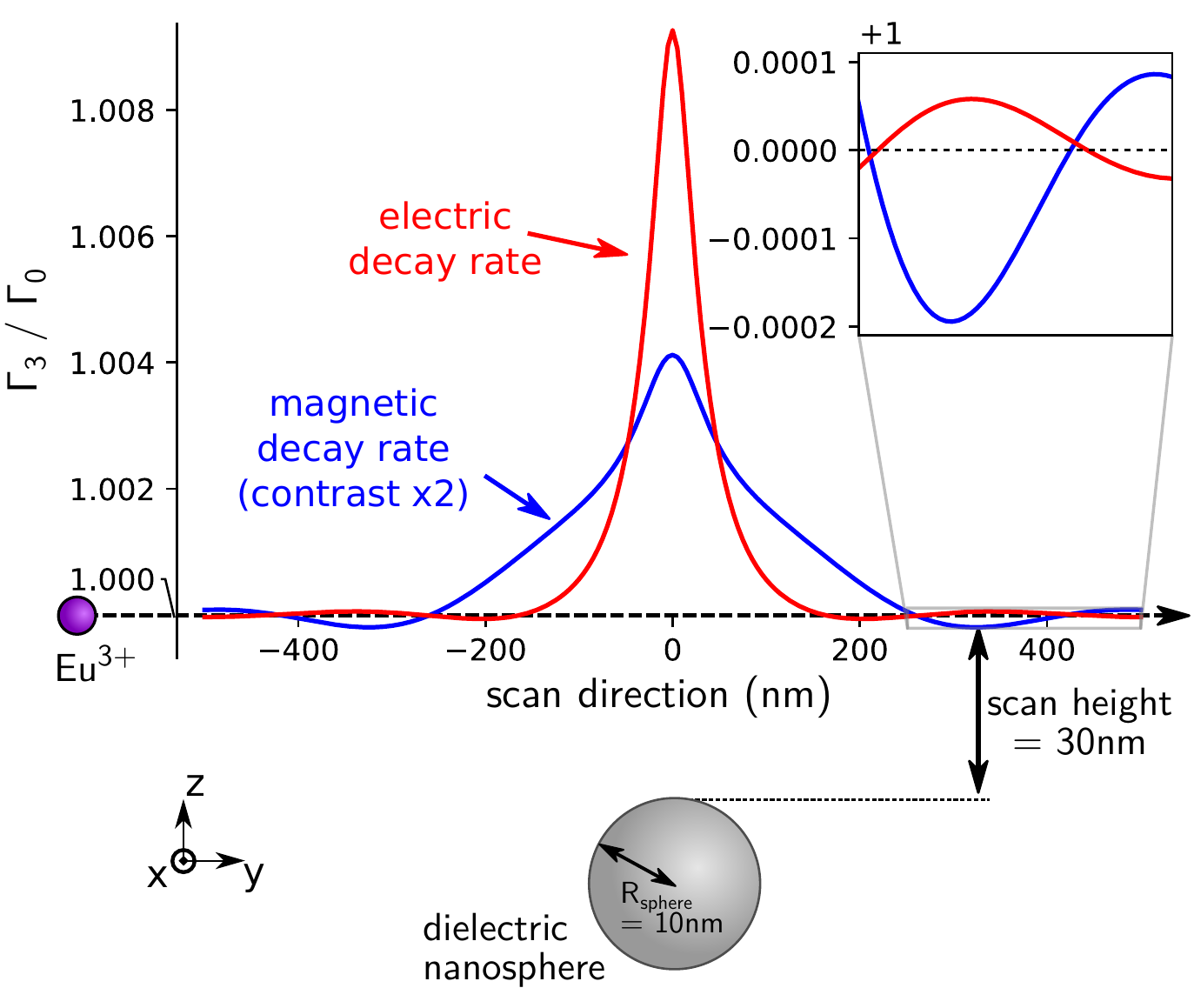}
	\caption{
		Decay rate of a single electric (red, \(\lambda_{\text{ED}}=610\,\)nm) and magnetic (blue, \(\lambda_{\text{MD}}=590\,\)nm) dipole transition (orientation averaged, relative to an isolated emitter), scanned at fixed height (\(30\)\,nm) above the surface of a silicon nano-sphere of radius \(R_{\text{sphere}}=10\,\)nm placed in vacuum. 
		The sphere lies at the center of the coordinate system.
	}
	\label{fig:nanosphere_ldos_E_M}
\end{figure}

As schematically depicted in Fig.~\ref{fig:nanosphere_ldos_E_M}, we consider a SNOM tip with a single Eu\(^{3+}\) emitter, which is raster-scanned along a line above a very small nano-particle placed in vacuum. The system is illuminated by a plane wave of wavelength \(\lambda_0=532\,\)nm at normal incidence and with linear polarization (along \(OY\)). 
We will also assume for the moment that the SNOM tip is interacting only weakly with the plane wave, so we neglect its optical scattering.

While our formalism allows to treat emitters of specific orientation without further modification (using the partial LDOS, e.g. Eq.~(\ref{eq:17}) for the magnetic case. See also Appendix~\ref{appendix:dipole_orientations}), in practical cases the SNOM tip is usually doped by a large number of rare earth emitters, where effectively an average of many orientations is recorded.
The orientation-averaged magnetic LDOS $n^{H}(\mathbf{r},\omega)$, influencing the MD transition, writes:
\begin{equation}\label{rhoH}
n^{H}(\mathbf{r},\omega) = n^{(0)}(\mathbf{r},\omega)+\dfrac{1}{2\pi^{2}\omega}\text{Im}\left[\Tr(\mathbf{S}_{p}^{HH}(\mathbf{r},\mathbf{r},\omega))\right]
\end{equation}
where the field susceptibility $\mathbf{S}_{p}^{HH}(\mathbf{r},\mathbf{r},\omega)$ is defined as~\cite{wiecha_decay_2018}:
\begin{equation}\label{eq:SpHH_definition}
\mathbf{S}_{p}^{HH}(\mathbf{r}, \mathbf{r}, \omega)=\alpha(\omega)\mathbf{S}_{0}^{HE}(\mathbf{r}, 0, \omega)\cdot\mathbf{S}_{0}^{EH}(0, \mathbf{r}, \omega)\, .
\end{equation}
Here \(\alpha\) is the polarizability of the observed nano-sphere, \(\omega\) the angular frequency of the illumination and $R_{\text{sphere}}$ its radius \cite{draine_discrete-dipole_1988}
\begin{equation}\label{eq:alpha_calusius_mossotti}
\alpha(\omega)=R_{\text{sphere}}^{3}\dfrac{\epsilon(\omega)-1}{\epsilon(\omega)+2} \, .
\end{equation}
The mixed field susceptibility is \cite{sersic_magnetoelectric_2011, schroter_modelling_2003, girard_optical_1997}
\begin{equation}
\mathbf{S}_{0}^{EH}(\mathbf{r}, \mathbf{r}', \omega)=ik_{0}\nabla\wedge \mathbf{G}_{0}(\mathbf{r}, \mathbf{r}', \omega)=\mathbf{S}_{0}^{HE}(\mathbf{r}', \mathbf{r}, \omega) \, ,
\end{equation}
where $\mathbf{G}_{0}(\mathbf{r}, \mathbf{r}', \omega)$ is the vacuum Green's tensor (e.g. see Refs. \cite{agarwal_quantum_1975} or \cite{wiecha_decay_2018}). 
Developing the tensor product in \eqref{eq:SpHH_definition} while assuming an emitter in the plane $\mathbf{r}_{0}=(0,y_{0},z_{0})$ simplifies the expression for \(\mathbf{S}_{p}^{HH}\) such that the different contributions to the \textit{magnetic} LDOS in the near- and far-field become easily discernible:
\begin{equation}\label{eq:SpHH_explicit_simplified}
\mathbf{S}_{p}^{HH}(\mathbf{r}_{0}, \mathbf{r}_{0}, \omega)=\alpha(\omega)C e^{2ik_{0}r_{0}}\left\lbrace -\dfrac{k_{0}^{4}}{r_{0}^{4}}-\dfrac{2ik_{0}^{3}}{r_{0}^{5}}+\dfrac{k_{0}^{2}}{r_{0}^{6}}\right\rbrace \, .
\end{equation}
Where the matrix $C$ is
\begin{equation}
C =
\begin{pmatrix}
z_{0}^{2}+y_{0}^{2} & 0 & 0 \\
0 & z_{0}^{2} & -z_{0}y_{0} \\
0 & -z_{0}y_{0} & y_{0}^{2}
\end{pmatrix}\, .
\end{equation}
We can do the same with the orientation-averaged electric LDOS 
\begin{equation}\label{rhoE}
n^{E}(\mathbf{r},\omega) = n^{(0)}(\mathbf{r},\omega) + \dfrac{1}{2\pi^{2}\omega}\text{Im}\left[\Tr(\mathbf{S}_{p}^{EE}(\mathbf{r},\mathbf{r},\omega))\right] \, .
\end{equation}
The \textit{electric-electric} field susceptibility $\mathbf{S}_{p}^{EE}(\mathbf{r},\mathbf{r},\omega)$ is given by
\begin{equation}\label{eq:SpEE_definition}
\mathbf{S}_{p}^{EE}(\mathbf{r},\mathbf{r},\omega)=\alpha(\omega)\mathbf{S}_{0}^{EE}(\mathbf{r}, 0, \omega)\cdot\mathbf{S}_{0}^{EE}(0, \mathbf{r}, \omega)
\end{equation}
where, using the notation of \cite{girard_near_2005}, the vacuum field susceptibility writes
\begin{multline}
\mathbf{S}_{0}^{EE}(\mathbf{r}, \mathbf{r}', \omega)= \\
\Big[-k_{0}^{2}\mathbf{T}_{1}(\mathbf{r}-\mathbf{r}')-ik_{0}\mathbf{T}_{2}(\mathbf{r}-\mathbf{r}')+\mathbf{T}_{3}(\mathbf{r}-\mathbf{r}')\Big] 
e^{ik_{0}\vert\mathbf{r}-\mathbf{r}'\vert}
\end{multline}
with
\begin{align}
\mathbf{T}_1(\mathbf{r}-\mathbf{r}') & = \dfrac{(\mathbf{r}-\mathbf{r}')(\mathbf{r}-\mathbf{r}') - \mathbf{I}\vert\mathbf{r}-\mathbf{r}' \vert^2}{\vert\mathbf{r}-\mathbf{r}' \vert^3} \\
\mathbf{T}_2(\mathbf{r}-\mathbf{r}') & = \dfrac{3(\mathbf{r}-\mathbf{r}')(\mathbf{r}-\mathbf{r}') - \mathbf{I}\vert\mathbf{r}-\mathbf{r}' \vert^2}{\vert\mathbf{r}-\mathbf{r}' \vert^4} \\
\mathbf{T}_3(\mathbf{r}-\mathbf{r}') & = \dfrac{3(\mathbf{r}-\mathbf{r}')(\mathbf{r}-\mathbf{r}') - \mathbf{I}\vert\mathbf{r}-\mathbf{r}' \vert^2}{\vert\mathbf{r}-\mathbf{r}' \vert^5} \, ,
\end{align}
where \(\mathbf{I}\) is the identity matrix. 
The three dyadic tensors \(\mathbf{T}_3\), \(\mathbf{T}_2\) and \(\mathbf{T}_1\) describe effects from the near- to the far-field. 
Explicitly, this gives
\begin{multline}\label{SpEE}
\mathbf{S}_{p}^{EE}(\mathbf{r},\mathbf{r},\omega) = \\ \alpha(\omega)e^{2ik_{0}r}
\Big[k_{0}^{4}\mathbf{T}_{1}\cdot\mathbf{T}_{1}+2ik_{0}^{3}\mathbf{T}_{1}\cdot\mathbf{T}_{2}-2k_{0}^{2}\mathbf{T}_{1}\cdot\mathbf{T}_{3}\\
-2ik_{0}\mathbf{T}_{2}\cdot\mathbf{T}_{3}-k_{0}^{2}\mathbf{T}_{2}\cdot\mathbf{T}_{2}+\mathbf{T}_{3}\cdot\mathbf{T}_{3} \Big] \, .
\end{multline}
By performing the tensor products and once more assuming a dipole at a location in the plane $\mathbf{r}_{0}=(0,y_{0},z_{0})$, we obtain an expression in which we can identify the different near- and far-field contributions to the \textit{electric} LDOS:
%
\begin{multline}\label{eq:SpEE_explicit_simplified}
\mathbf{S}_{p}^{EE}(\mathbf{r}_{0},\mathbf{r}_{0},\omega) =\\
\quad \alpha(\omega)e^{2ik_{0}r_{0}}\Bigg[
 \left\lbrace\dfrac{k_{0}^{4}}{r_{0}^{4}}+\dfrac{2ik_{0}^{3}}{r_{0}^{5}}-\dfrac{2k_{0}^{2}}{r_{0}^{6}} \right\rbrace A  + \left\lbrace \dfrac{1}{r_{0}^{8}}-\dfrac{2ik_{0}}{r_{0}^{7}}-\dfrac{k_{0}^{2}}{r_{0}^{6}} \right\rbrace B \Bigg]
\end{multline}
%
with 
\begin{equation}
A=
\begin{pmatrix}
z_{0}^{2}+y_{0}^{2} & 0 & 0 \\
0 & z_{0}^{2} & -y_{0}z_{0} \\
0 & -y_{0}z_{0} & y_{0}^{2}
\end{pmatrix}
\end{equation}
and
\begin{equation}
B=
\begin{pmatrix}
z_{0}^{2}+y_{0}^{2} & 0 & 0 \\
0 & 4y_{0}^{2}+z_{0}^{2} & 3y_{0}z_{0} \\
0 & 3y_{0}z_{0} & 4z_{0}^{2}+y_{0}^{2}
\end{pmatrix} \, .
\end{equation}

\begin{figure}[t]
	\centering
	\includegraphics[width=\columnwidth]{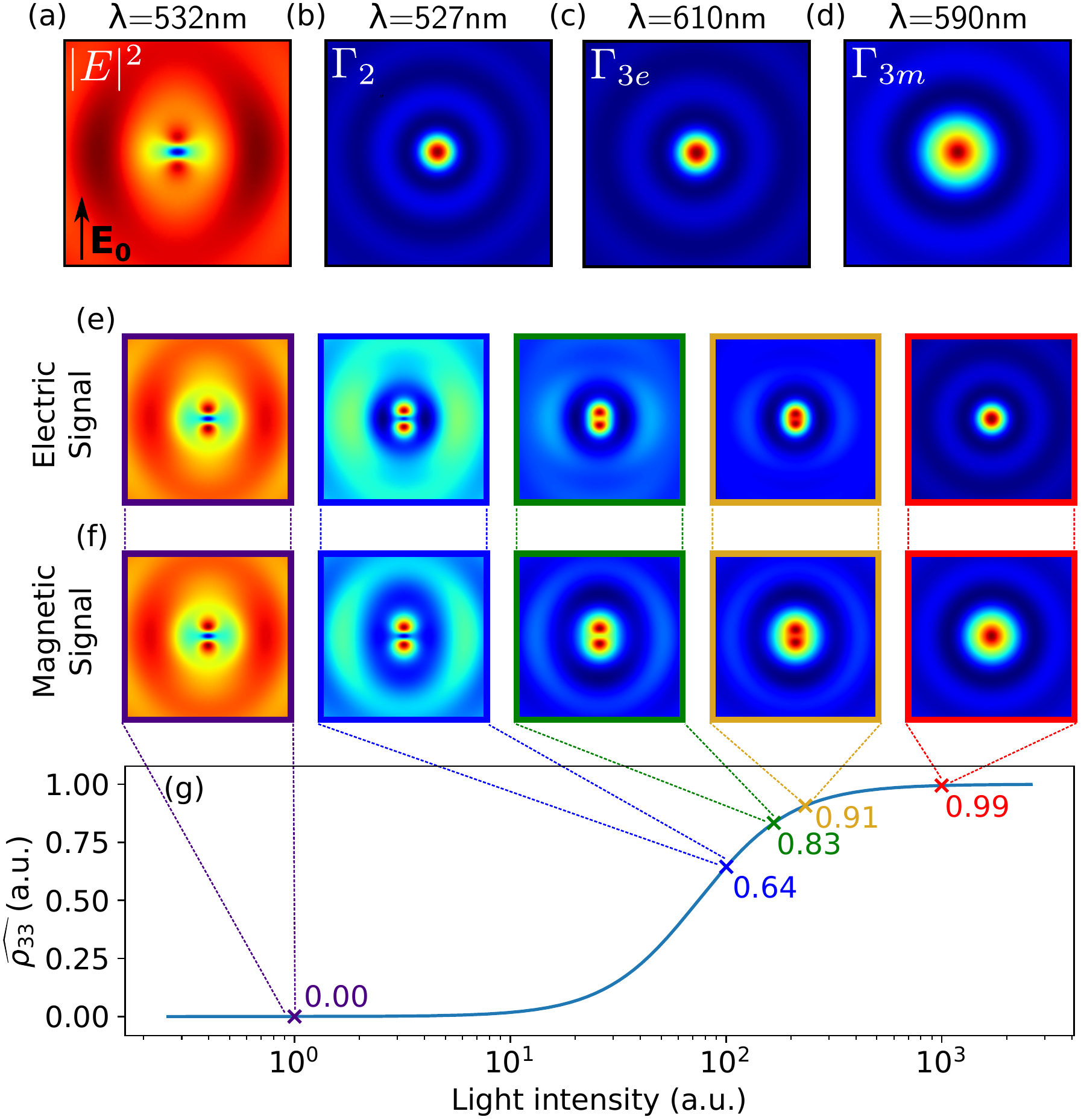}
	\caption{
		Evolution of the simulated electric and magnetic signal for increasing illumination intensity, calculated for a single europium atom, raster-scanned above a plane wave illuminated silicon nano-sphere (\(R_{\text{sphere}}=10\,\)nm) in vacuum. 
		The atom is raster-scanned at fixed heigh $\Delta z$=60nm above the top surface of the Si sphere.
		(a) Near field map at \(\lambda_0=532\,\)nm, (b)-(c) electric decay rate maps at $\lambda_{\text{ED}}=527\,$nm and $\lambda_{\text{ED}}=610\,$nm and (d) magnetic decay rate map at $\lambda_{\text{MD}}=590\,$nm in the plane defined by $\Delta z = 60\,$nm.
		(e) Electric and (f) magnetic europium signal maps, each map is related to the population specified with the same color of the marker as used for the frame around the map.
		(g) A semi-logarithmic representation of the saturation curve of the active level $\widehat{\rho_{33}}$ as a function of the laser intensity, when the atom is centered above the sphere.
		The maps show areas of \(1200\times 1200\,\)nm\(^2\).
	}
	\label{fig:saturation_transition}
\end{figure}

\begin{figure}[t!]
	\centering
	\includegraphics[width=.9\columnwidth]{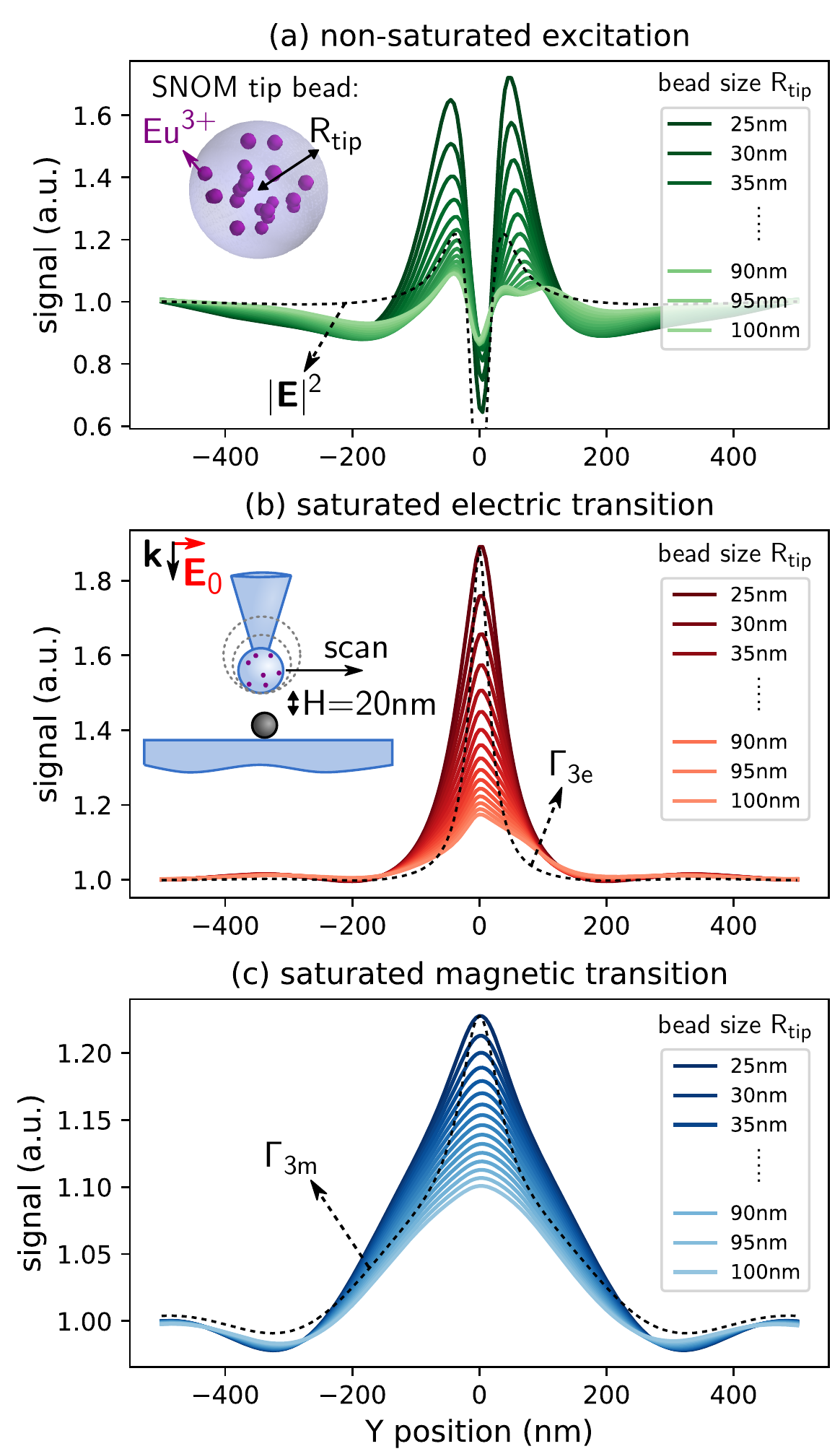}
	\caption{
		Influence of the SNOM tip size (increasing radius \(R_{\text{tip}}\) of Eu\(^{3+}\)-doped bead) on the recorded SNOM signal.
		The SNOM tip is doped by 20 randomly distributed Eu\(^{3+}\) emitters (see inset in (a)).
		As illustrated in the inset in (b), the tip is scanned above a spherical silicon particle (sphere of radius \(10\,\)nm), lying on a dielectric substrate (\(n_{\text{s}}=1.5\)) in air (\(n_{\text{env}}=1\)).
		The scanning plane height is fixed at \(H=20\,\)nm above the surface of the silicon sphere (distance tip bead bottom surface to Si sphere top surface).
		The figures show the signal of the electric transition \(\Gamma_{3e}\) under (a) non-saturated and (b) saturated excitation and of (c) the magnetic europium transition \(\Gamma_{3m}\) under saturation for increasing SNOM-bead size.
		Dashed black lines indicate (a) the near-field intensity (b) the electric decay rate \(\Gamma_{3e}\) and (c) the magnetic decay rate \(\Gamma_{3m}\) at the bottom surface of the probe, the data is re-normalized to the scale of the signal. 
		(b-c) are plotted on the same scale, the signal in (a) is several orders of magnitude weaker and normalized separately.
		This random arrangement of the emitters causes a slight asymmetry in the signal profiles.
	}
	\label{fig:tip_size}
\end{figure}

Figure~\ref{fig:nanosphere_ldos_E_M} shows the orientation-averaged electric (red) and magnetic (blue) decay rates above a silicon nano-sphere of \(10\,\)nm radius, raster-scanned by a single europium atom at a height of \(30\,\)nm.
We observe that, while at short distances to the nano-sphere, the electric and magnetic decay-rates have the same, positive sign.
At the transition to the far-field, an inversion in the contrast between \(\Gamma_e\) and \(\Gamma_m\) occurs \cite{lunnemann_local_2016}. 
This reflects the opposite signs in the \(r^{-4}\) and \(r^{-5}\) terms of equations~(\ref{eq:SpHH_explicit_simplified}) and~(\ref{eq:SpEE_explicit_simplified}).

\section{From a point-like europium emitter to a doped SNOM tip}

We now want to analyze the influence of the excitation light intensity (i.e. the Rabi frequency \(\Omega\)) on the SNOM signal.
We can use the calculated decay rates together with the self-consistent electric field at the location of the europium atoms to obtain the population \(\widehat{\rho_{33}}\) of the light-emitting Eu\(^{3+}\) energy level \(^5\)D\(_0\) and finally the signal Eq.~(\ref{eq:12}), emitted by the europium ion.
We determine these via a volume discretization~\cite{girard_near_2005}, implemented in our home-made python toolkit ``pyGDM''~\cite{wiecha_pygdmpython_2018}.
Since the rates depend on several factors like the embedding polymer \cite{shavaleev_influence_2015}, we choose for the decay rates $\Gamma_{3e/m}^{(0)}$ of the meta-stable electric and magnetic energy levels \(\vert 3\rangle\) a value of \(1\,\)ms\(^{-1}\).
Relative to the \(\Gamma_{3e/m}\) transitions, we chose the \(\Gamma_2\) transition to be \(\times 10^3\) faster and the K transition to decay an additional order of magnitude faster (see table~\ref{tab:decay_rates_sim}).
We emphasize that this is an arbitrary choice of the decay rates for our numerical demonstration.
In simulations aiming at the comparison to measurements, experimentally determined values for the decay rates of the employed probe should be used.

\begin{table}
	\begin{center}
		\begin{tabular}{l l l}
			\toprule
			transition & corresponding levels & decay rate (ms$^{-1}$) \\[0.1cm]
			\midrule 
			$\Gamma_{2}^{(0)}$ & $^{5}D_{1}\rightarrow$ $^{7}F_{1}/^{7}F_{2}$ & 1000 \\ 
			K$^{(0)}$ & $^{5}D_{1}\rightarrow ^{5}D_{0}$ & 10000 \\
			$\Gamma_{3m/e}^{(0)}$ & $^{5}D_{0}\rightarrow$ $^{7}F_{1}/^{7}F_{2}$ & 1\\
			\bottomrule
		\end{tabular}
	\end{center}
	\caption{
		Decay rates of the considered transitions in our numerical examples. The first column represents the transition, the second gives its initial and final energy level. 
		The decay rate of the isolated emitter (without nanostructure) is given in the third column.
	}
	\label{tab:decay_rates_sim}
\end{table}

\subsection{Point-like, single europium emitter}

We show in figure~\ref{fig:saturation_transition}(a) the electric field intensity in a plane at a distance \(\Delta z = 60\,\)nm above the surface of the \(R_{\text{sphere}}=10\,\)nm silicon nano-sphere illuminated by a plane wave of wavelength \(\lambda_0=532\,\)nm under normal incidence (linear polarization along \(Y\)).
The maps \ref{fig:saturation_transition}(b-d) show maps of the (orientation averaged) decay rates \(\Gamma_2\), \(\Gamma_{3e}\) and \(\Gamma_{3m}\).

We use these quantities to calculate after equation~(\ref{eq:12}) the signal of the europium atom at every position of the scanned area.
The signal due to the electric and due to the magnetic transition is plotted in Fig.~\ref{fig:saturation_transition}(e), respectively \ref{fig:saturation_transition}(f) for increasing intensities of the illuminating light. 
The illumination intensities are indicated by colored markers in figure~\ref{fig:saturation_transition}(g), which shows the population \(\widehat{\rho_{33}}\) of the emitting energy level as function of the excitation intensity.
For the calculation of \(\widehat{\rho_{33}}\) in \ref{fig:saturation_transition}(g), the europium atom is centered on top of the silicon sphere.
We see that in the case of an illumination below saturation (purple marker in \ref{fig:saturation_transition}(g)), both, the MD as well as the ED transition signals correspond to the near-field intensity distribution.
On the other hand, if the energy-level \(\vert 3 \rangle\) is saturated due to strong illumination intensity, the recorded signal is proportional to the local decay rate \(\Gamma_{3e/m}\), hence to the relative electric (using the ED transition) or magnetic (using the MD transition) LDOS.
For intermediate light intensities, the maps undergo a transition between near-field distribution and LDOS. 
We note, that almost full saturation (\(\gtrsim 0.9\) in Fig.~\ref{fig:saturation_transition}(g)) is required to yield a signal proportional to the LDOS.

\subsection{Large probes with inhomogeneous emitter density}

\begin{figure*}[t!]
	\centering
	\includegraphics[width=.9\textwidth]{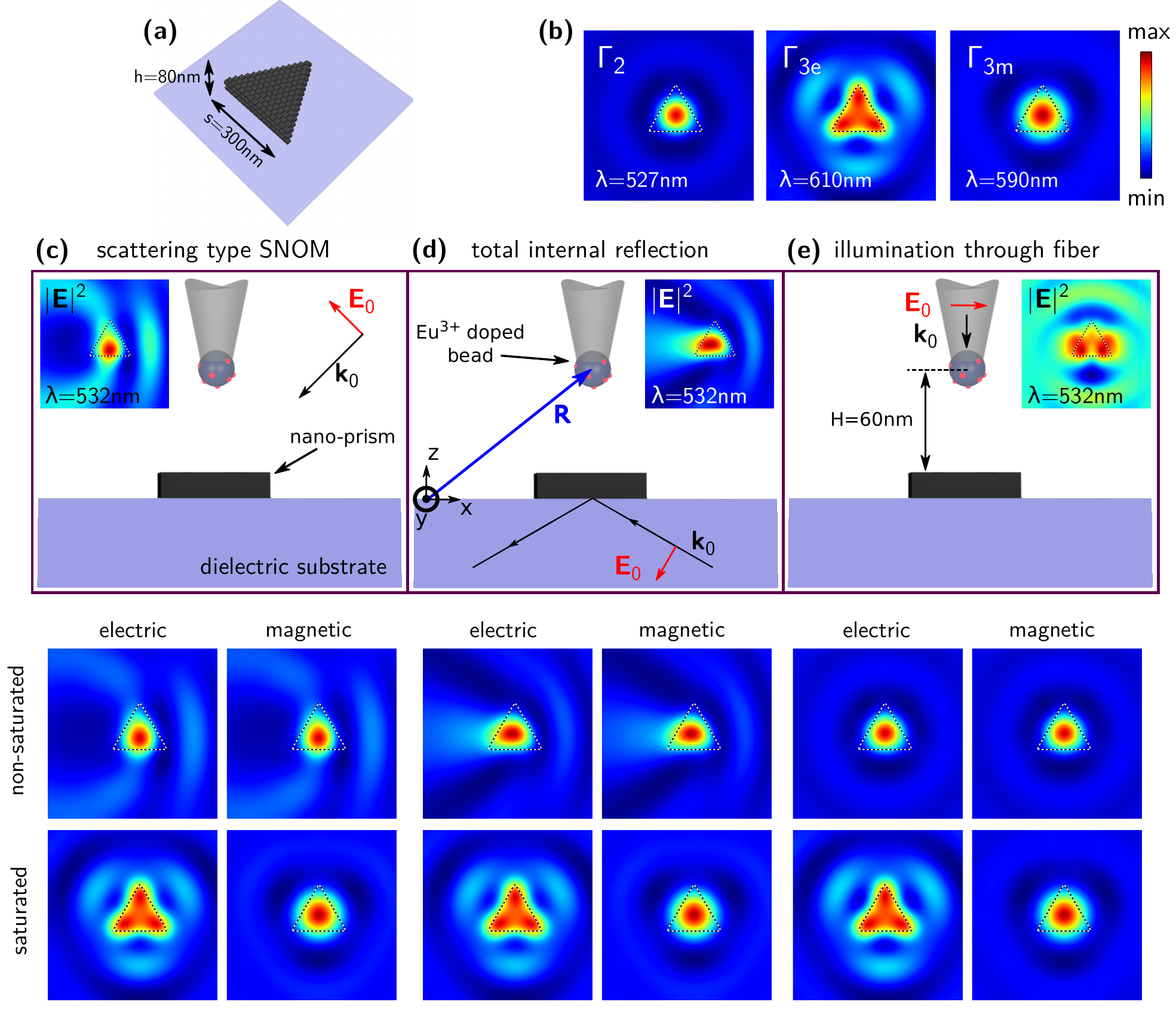}
	\caption{
		SNOM signal from raster-scanning a \(R_{\text{tip}}=50\,\)nm large Eu\(^{3+}\) doped nano-bead above a equilateral triangular silicon nano-prism of uniform side lengths \(s=300\,\)nm and height \(h=80\,\)nm, lying on a dielectric substrate (\(n_{\text{s}}=1.5\)) surrounded by air. 
		The scanning plane is parallel to the substrate, the center of the SNOM bead being at a height \(H=60\,\)nm above the nanoprism's top surface.
		(a)~The geometry of the nano-prism and its dimensions.
		(b)~scanning plane mappings of the decay rates that contribute to the SNOM image formation. 
		(c-e) Different SNOM configurations. 
		Top subplots: illustration of the considered optical setup. 
		Bottom plots: signal from the europium ED (left column) and MD (right column) transition under non-saturated (top row) and saturated (bottom row) excitation.
		(c)~scattering type SNOM illuminated with a \textit{p}-polarized plane wave at \(45^{\circ}\) angle of incidence. 
		(d)~total internal reflection SNOM, illuminated with a \textit{p}-polarized plane wave from below at the critical angle (\(-45^{\circ}\)). 
		(e)~fiber-illuminated (apertured) SNOM. Linear (\(x\)-) polarized illumination of the bead through a fiber.
		Incident wavelength is \(\lambda_0=532\,\)nm. All color-maps show areas of~\(1\times1\,\)\textmu m\(^2\).
	}
	\label{fig:different_snom_configs}
\end{figure*}

So far we considered a single europium emitter as SNOM probe. However, in realistic cases active SNOM probes are usually rather large, polymer-based or other dielectric nano-particles with diameters of at least around \(100\,\)nm \cite{michaelis_optical_2000, aigouy_mapping_2014, cuche_fluorescent_2009, cuche_near-field_2017, sanz-paz_enhancing_2018, ernandes_exploring_2018}.
We therefore use our method in the following, to assess the impact of a larger detection volume, in which emitters might be dispersed with an inhomogeneous density distribution.
To simulate such inhomogeneous emitter densities, we randomly distribute~$N_{\text{q}}=20$ europium atoms in a spherical volume of radius~\(R_{\text{tip}}\), which represents the emitter-doped bead at the SNOM tip apex.
In the inset of figure~\ref{fig:tip_size}(a) we show a sketch of this numerical model.
We now rasterscan the assembly of~\(20\) europium atoms on a line at constant height defined by a fixed spacing of \(H=20\,\)nm between the SNOM tip and the nano-object, as illustrated in the inset of figure~\ref{fig:tip_size}(b). 
The imaged object is again a silicon nano-sphere of radius \(R_{\text{sphere}}=10\)\,nm.
At each tip position we calculate the signal of the ensemble of emitters via equation~(\ref{eq:13}).

For non-saturated illumination (figure~\ref{fig:tip_size}(a)), we found earlier that the signal of each Eu\(^{3+}\) ion is proportional to the local near-field intensity. 
Since we now calculate the sum of signals from different locations, we expect to obtain some broadening if the SNOM tip increases in size. 
Indeed, while for a small active volume the spatial near-field distribution is reproduced quite accurately, the signal from larger tip beads broadens and the intensity decreases (light green lines in figure~\ref{fig:tip_size}(a)). The decrease of signal is due to the fixed amount of europium emitters in our model. To reflect constant emitter density, this could be multiplied by a volume-scaling factor, which however is not the main point here.
More importantly, since some emitters of the random Eu\(^{3+}\) distribution are now much closer to the probed object than others, the closest near-field zone with usually strongest intensity enhancement is probed by only a few, randomly distributed europium atoms. 
The random distribution finally leads to some non-symmetric features in the recorded signal.

In case of saturation, shown in figure~\ref{fig:tip_size}(b-c), an increase of the SNOM bead size leads again to a broadening of the signal, which can be observed in case of the electric (\ref{fig:tip_size}(b)) as well as magnetic (\ref{fig:tip_size}(c)) transition. 
The ED signal is proportional to \(\Gamma_{3e}\), which we found to strongly increase in the close proximity of a nano-structure due to the near-field term \(\mathbf{T}_3\) (see equation~(\ref{eq:SpEE_explicit_simplified}) and Fig.~\ref{fig:nanosphere_ldos_E_M}). 
In consequence, in a large SNOM tip, few emitters in the SNOM bead which are closest to the probed nanostructure, now contribute particularly strongly to the full signal. 
The random distribution of emitters furthermore leads to an offset in the peak intensity and to an asymmetric line-shape.

In summary, while we find that SNOM tip volumes of important size also lead to signals that are proportional to the nearfield intensity (under non-saturated excitation) or to the electric or magnetic LDOS (in the saturated regime), spatial features are broadened with increasing probe size, reducing the spatial resolution of the SNOM. 
Furthermore, inhomogeneous emitter distributions in the active volume can distort the signals and may lead to asymmetric and shifted features in the spatial signal profiles.
In conclusion, as active SNOM probes are usually of relatively large size, a careful interpretation is crucial in the analysis of according experimental data. 
On the other hand, using specifically designed calibration samples, the above described distortion effects may to some extent be taken into account in data post-processing.

\subsection{Several typical SNOM configurations}

Finally, we want to compare and analyze the expected signal in several typical realistic SNOM configurations.
In this analysis we assume again a single point-like emitter in the SNOM bead. 
Of course the above findings for large probes apply here as well.
To furthermore demonstrate the capability of our formalism to calculate signals also for non-spherical nanostructures, we use for our demonstration a equilateral triangular silicon nano-prism, as depicted in figure~\ref{fig:different_snom_configs}(a).
Thanks to the concept of a generalized propagator \cite{martin_generalized_1995, girard_near_2005}, also in case of such arbitrarily shaped particles, the e-/m-LDOS as well as the total field intensity at the location of the europium atom can be calculated very efficiently at the different probe-positions~\cite{teulle_scanning_2012}.
The mappings of the optical transition decay rates in the scanning plane are depicted in Fig.~\ref{fig:different_snom_configs}(b). 
The illumination electric field intensity distributions for the respective SNOM configurations are shown in the insets in the SNOM schematics at the top of Figs.~\ref{fig:different_snom_configs}(c-d).

Figure~\ref{fig:different_snom_configs}(c) shows the case of a scattering-type (``apertureless'') SNOM, which corresponds essentially to the configuration that we considered until now. 
The only difference is that we here consider an illumination at \(45^{\circ}\) oblique angle of incidence (shown: ``\(p\)'' polarization).
As expected, at low intensities (top panels just below the illustration in figure~\ref{fig:different_snom_configs}(c)) the signal of both ED and MD transitions correspond to the electric field intensity distribution. 
In the saturated scenario on the other hand, according to the observed quantum transition we obtain a mapping of either the electric or the magnetic LDOS.

In another typical type of SNOM, the illumination is established via total internal reflection, in which case the sample is deposited on a dielectric prism, whose surface is illuminated from below at critical angle.
This is schematically shown at the top of figure~\ref{fig:different_snom_configs}(d). 
If the illumination is kept at a low intensity, again the near-field intensity distribution is detected by the raster-scanned SNOM probe (c.f. inset in illustration). 
Likewise, when the energy level \(\vert 3 \rangle\) is saturated, a map of the part of the LDOS is obtained, that corresponds to the observed optical transition.
The main difference is here, that the excitation is now due to an evanescent field at the substrate interface, leading to a different near-field distribution at the sample.

Finally, we consider the case of a fiber-illuminated SNOM (figure~\ref{fig:different_snom_configs}(e)).
Here the Eu\(^{3+}\) doped nano-bead is attached to an optical fiber through which it is illuminated by a laser.
We model the illumination by a plane wave at the position of the nano-bead, but which elsewhere is zero. 
Only indirectly, a small portion of the laser light reaches the sample after scattering at the SNOM tip. 
In the case of high laser intensity, hence when saturating the emitters, this configuration is again capable to detect the electric and magnetic parts of the photonic LDOS. 
However, at low intensities where the other configurations are sensitive to the local field, here the situation is more complex. 
As mentioned above, the nanostructure is illuminated by the scattered laser-light at the SNOM apex nano-bead. 
This corresponds to illuminating the system by a (weak) dipolar source. 
The total field intensity at the nano-bead location (inset in the illustration in figure~\ref{fig:different_snom_configs}(e)) is now a superposition of the laser illumination arriving through the fiber, and the comparatively very weak back-scattering by the nanostructure. 
Compared to the back-scattering from the nanostructure, the contribution of the laser is strong, hence the field variations due to scattering at the nanostructure are very weak. 
In consequence, other contributions to the population density of the emitting energy level Eq.~(\ref{eq:10}) may become visible.
Indeed, here we obtain a mapping which resembles the LDOS at the wavelength of the excitation transition \(\Gamma_2\).
In our model the transition \(\vert 1 \rangle \rightarrow \vert 2 \rangle\) is of electric dipolar  nature in both the ED and the MD emission case. 
Hence, the low-intensity signal is identical for the electric and the magnetic transition.
We note that this behavior is dependent on the relative decay rates in the 3-level quantum system.
For instance, in an emitter system with much higher decay rate \(\Gamma_2\), the signal under non-saturated excitation corresponds again to the electric field intensity distribution, rather than to \(\Gamma_2\).
When interpreting experimental data of such a SNOM system it is therefore important to carefully choose the relative decay rates of the employed quantum system.

We finally note that the signal here is entirely shaped by the near-field interaction and photo-dynamics in the electronic 3-level system of the europium. 
While in SNOM configurations with passive probe, a large solid angle of detection is required to get access to the photonic LDOS \cite{colas_des_francs_relationship_2001}, this limitation does not affect active SNOM probes, which facilitates the implementation of the detection scheme in according experiments.

\section{Conclusions}

In conclusion, we presented a quantum theory able to fully describe the optical interactions and photo-dynamics in scanning optical near-field microscopy using rare-earth ions as active quantum probes. 
Such SNOM configuration allows to probe separately the \textit{electric} and the \textit{magnetic} parts of the photonic LDOS. 
Our approach allows not only to theoretically model the SNOM image formation associated to these two parts of the LDOS.
The theory is furthermore capable to fully describe the transition from the non-saturated to the saturated excitation regime.
We find that while low excitation intensity generally results in SNOM signals proportional to the electric near-field intensity, saturated emitters yield a signal corresponding to the electric or magnetic LDOS, depending on the detected quantum transition (ED or MD).
Together with a numerical discretization scheme our method can be applied to nanostructures of arbitrary shape and material and a substrate can be included without any effort. 
Moreover we demonstrated that the method can be easily adapted to various SNOM configurations.
We foresee that using experimental constants for the energy levels and decay rates of the concerned quantum transitions, our theory will be very useful to interpret active-probe SNOM images and to estimate in which regime (saturation or not) the SNOM data is recorded. 
Our model can be directly adapted to the case of resonant driving of a magnetic transition at the excitation using focused vector beams~\cite{kasperczyk_excitation_2015}, in which case an active SNOM probe might be rendered sensitive also to the \textit{intensity distribution} of the optical magnetic field.
Finally, the approach is not limited to rare-earth elements and can be applied to other types of quantum emitter systems.


\section{Appendix}

\subsection{Emitter orientation dependence of SNOM signal}\label{appendix:dipole_orientations}

\begin{figure}[t!]
	\centering
	\includegraphics[width=\columnwidth]{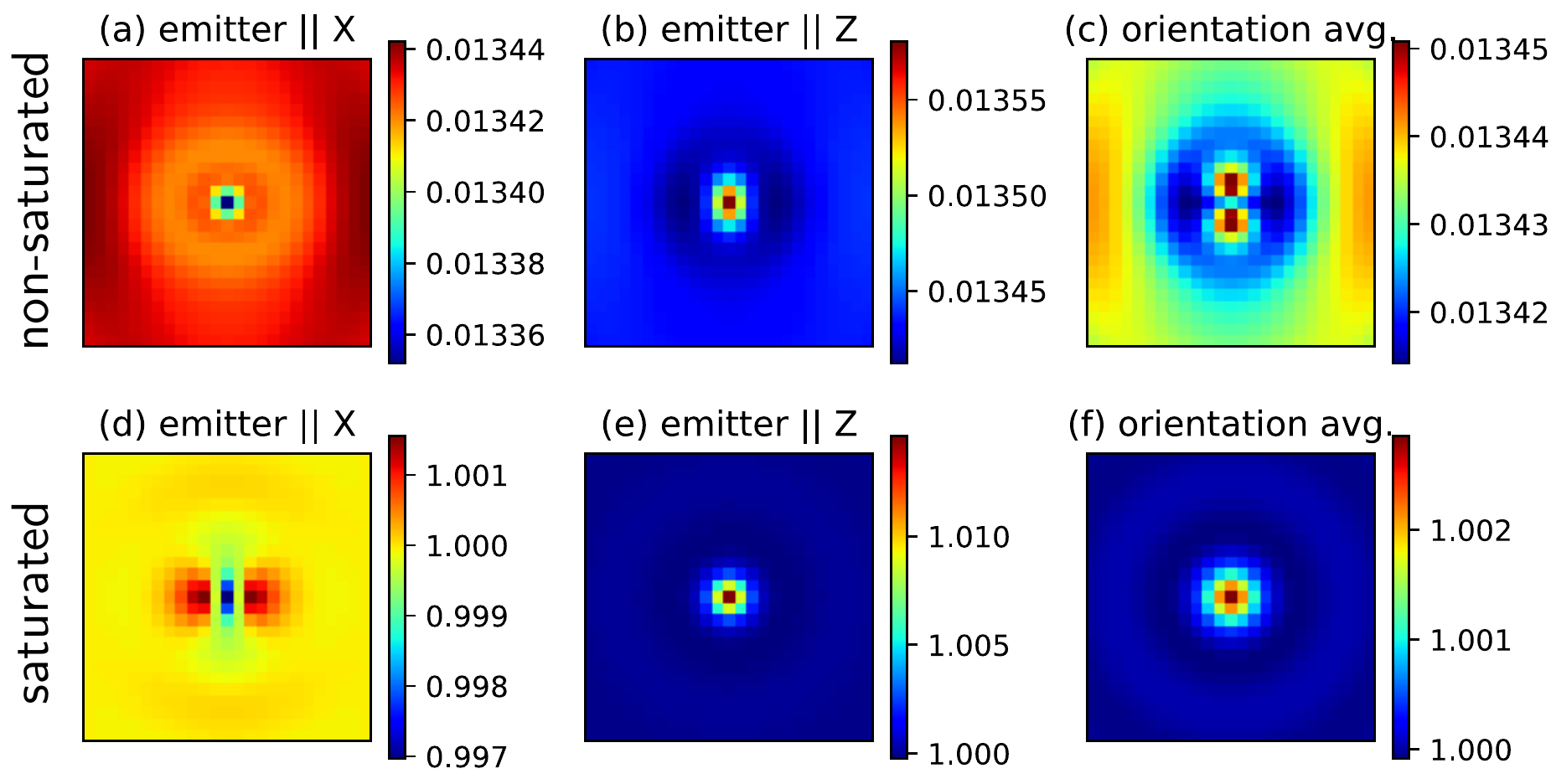}
	\caption{
		Same simulation as in figure~\ref{fig:saturation_transition}e, but using a dipole of specific orientation. 
		(a) and (d) electric dipole signal for an emitter along $X$.
		(b) and (e) ED transition along $Z$.
		(c) and (f) ED transition orientation averaged, hence same situation as in figure~\ref{fig:saturation_transition}e, left-most and right-most panels.
		Normally incident plane wave with \(\lambda_0=532\,\)nm, emitter atom scanned in a plane at \(\Delta z=60\,\)nm above the top surface of a silicon sphere of \(10\,\)nm radius, lying on an \(n_{\text{s}}=1.45\) substrate.
		(a-c) corresponds to non-saturated, (d-f) to saturated excitation.
		Color scale corresponds to the SNOM emitter signal in arbitrary units.
		Shown areas are \(800\times 800\)\,nm\(^2\).
	}
	\label{fig:appendix_emitter_orientation}
\end{figure}

In practical situations, a europium SNOM tip contains usually a large number of rare earth atoms, thus the average orientation of the quantum emitters is arbitrary. 
To simulate such situation, we take the trace of the field susceptibility (Eqs.~(\ref{rhoH}) and~(\ref{rhoE})). 
Our formalism is however not restricted to using the orientation-average of the emitters. 
By using the \textit{partial} LDOS (see Eq.~(\ref{eq:17})) instead of the trace, we obtain the SNOM signal due to an emitter of a specific orientation. 
In cases where a preferential orientation of the dipole moments occurs, it can have a significant impact on the recorded signal, which is demonstrated in figure~\ref{fig:appendix_emitter_orientation} at the example of the SNOM signal due to an electric dipole transition.
The top row (\ref{fig:appendix_emitter_orientation}a-c) shows the case of non-saturated excitation of a single quantum emitter, raster-scanned on top of a small dielectric sphere.
Figure~\ref{fig:appendix_emitter_orientation}(d-f) show the same raster-scan in case of saturated excitation. 
The geometry and illumination conditions are identical to figure~\ref{fig:saturation_transition}.
Non-saturated and saturated excitation correspond to light intensities \(10^0\) and \(10^3\) (a.u.) in Fig.~\ref{fig:saturation_transition}, respectively.

\subsection{Influence of large SNOM tip on recorded SNOM signal}\label{appendix:LDOS_inside_snom_sphere}

\begin{figure}[t!]
	\centering
	\includegraphics[width=0.9\columnwidth]{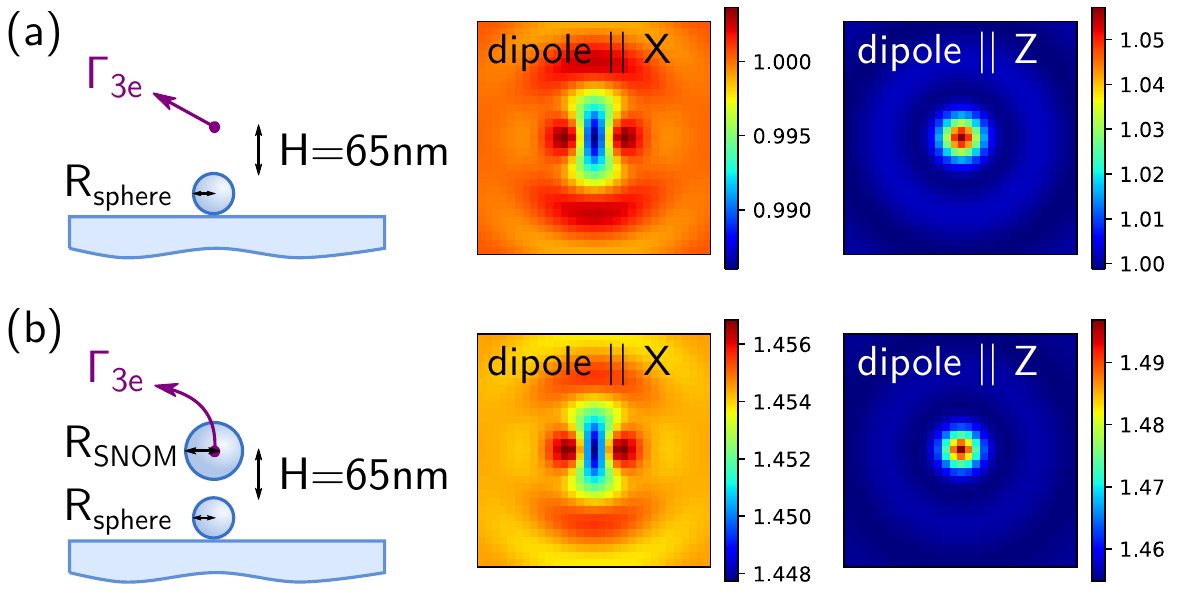}
	\caption{
		Relative electric decay rate at \(\lambda_{\text{ED}}=527\,\)nm calculated at \(H=65\,\)nm above the top surface of a small dielectric sphere of radius \(R_{\text{sphere}}=30\,\)nm (a) in vacuum and (b) at the center inside an actually simulated dielectric sphere, representing the doped SNOM tip (\(R_{\text{SNOM}}=50\,\)nm).
		Material of sphere and SNOM tip is identical (\(n_{\text{sphere}}=2\)), the sphere is lying on a glass substrate (\(n_{\text{s}}=1.45\)). 
		Shown areas are \(800\times 800\)\,nm\(^2\).
		The slight color difference in \(X\)-orientation is mainly due to a small shift in the color-range auto-scaling of our plotting software.
	}
	\label{fig:appendix_LDOS_inside_SNOM_tip_sphere}
\end{figure}

The LDOS inside a (dielectric) nano-sphere depends strongly on the size and supported modes of the particle \cite{schnieppSpontaneousEmissionEuropium2002}.
In the main text we approximated the optical contribution of the doped SNOM-tip material by a point-dipole model. 
In order to justify this approximation, we want to assess the impact of an actual sphere on the photonic LDOS at the location of the probe emitters.
To this end, we show in figure~\ref{fig:appendix_LDOS_inside_SNOM_tip_sphere}(a) the LDOS at a height \(H=65\,\)nm above the top surface of a small dielectric sphere (\(n_{\text{sphere}}=2\), \(R_{\text{sphere}}=30\,\)nm), deposited on a glass substrate (\(n_{\text{s}}=1.45\)). 
This situation corresponds to the model used throughout this article, where the SNOM tip was approximated as a point-dipole. 
This approximation takes into account only scattering of the illumination, but is neglecting the SNOM tip's influence on the relative LDOS at the emitter position.
In figure~\ref{fig:appendix_LDOS_inside_SNOM_tip_sphere}(b) we show a raster-scan simulation, in which the SNOM sphere was modelled with the same discretization as the probed object. 
The SNOM sphere is of the same material (\(n_{\text{SNOM}}=2\)), and discretized with the same step-size (\(s=15\,\)nm) but of larger size (\(R_{\text{SNOM}}=50\,\)nm). 
We calculate the LDOS \textit{inside} the discretized SNOM-tip sphere at the location of the centre mesh cell as described in section~4.3.2 of Ref.~\cite{wiecha_pygdmpython_2018}.
We note that we consider here an extreme case, where the SNOM tip is larger than the probed object. 
We find that, except of a global offset in the decay rate and a small scaling correction, the images in both cases (isolated emitter, emitter inside SNOM tip) are actually quasi identical.
Our model allows to incorporate the LDOS offset due to the material of the SNOM tip in the \(\Gamma_0\) values. 
In conclusion we find that the presence of the SNOM tip can be included in good approximation by a point-dipole model.

\bigskip
\noindent
{\bf Disclosures}:
The authors declare no conflicts of interest.
\par
\noindent

\bigskip
\noindent
{\bf Acknowledgments}:
The authors thank G. Colas des Francs for fruitful discussions.
PRW acknowledges support by the German Research Foundation (DFG) through a research fellowship (WI 5261/1-1).
AC acknowledges funding from the French Programme Investissements d’Avenir (ANR-11-IDEX-0002-02, reference ANR-10-LABX-0037-NEXT).
This work was supported by the computing center CALMIP in Toulouse.
All data supporting this study are openly available from the University of Southampton repository (DOI: 10.5258/SOTON/D1176).
\par
\bigskip
\noindent

\bibliography{paper-josaB-snom.bbl}



\end{document}